\documentclass[a4paper,12pt]{iopart}

\usepackage{amssymb}
\usepackage{graphicx}
\usepackage{color}
\usepackage{bm}
\begin{document}

\title{Exponential localization in one-dimensional quasiperiodic optical lattices}
\author{Michele Modugno}
\address{LENS \& Dipartimento di Fisica, University of Florence, Via N. Carrara 1 500019 Sesto Fiorentino, Italy}
\ead{modugno@fi.infn.it}

\begin{abstract}
We investigate the localization properties of a one-dimensional bichromatic optical lattice in the tight binding regime, by discussing how exponentially localized states emerge upon changing the degree of commensurability.  We also review the mapping onto the discrete Aubry-Andr\'e model, and provide evidences on how the momentum distribution gets modified in the crossover from extended to exponentially localized states.
This analysis is relevant to the recent experiment on Anderson localization of a noninteracting Bose-Einstein condensate in a quasiperiodic optical lattice [G. Roati \textit{et al.}, Nature \textbf{453}, 895 (2008)].
\end{abstract}
\maketitle

\section{Introduction}
One-dimensional quasiperiodic potentials have been the object of an extensive theoretical investigation since the 80's 
\cite{harper,aubry-andre,thouless,grempel,kohmoto,diophantine,diener,ingold,aulbach,holthaus}, 
owing to their interesting feature of having a spatial ordering that is intermediate between periodicity and disorder.
The paradigm of this class of models is the so called Harper \cite{harper} or Aubry-Andre model \cite{aubry-andre}, described by the following hamiltonian
\begin{equation}
H=-J\sum_{j} \left(c_{j+1}^*c_j + c_j^*c_{j+1}\right) + \Delta\sum_j\cos(2\pi \beta j+ \phi) |c_j|^2
\label{eq:aa-hamiltonian}
\end{equation}
that, for a certain class of irrational values of $\beta$ \cite{kohmoto,diophantine}, displays a ``metal-insulator'' phase transition from extended to exponentially localized states, just as Anderson localization in random systems in higher dimensions \cite{aubry-andre,anderson,kramer}. This model has the interesting properties of being self-dual under a suitable transformation to momentum space, the self-dual point being $\Delta/J=2$ at which the transition occurs \cite{aubry-andre}.

This kind of hamiltonians can now be engineered with ultracold atoms loaded in optical lattices \cite{damski,roth,lye,fallani,guarrera,roati}, opening the way to study the localization properties of quasiperiodic systems from the experimental point of view. The first experimental observation of Anderson localization in quasiperiodic potentials has been recently reported in \cite{roati,billy}. This experiment has been realized by using a non interacting Bose-Einstein condensate loaded in a bichromatic potential, obtained by superimposing two one-dimensional optical lattices with different wavelengths.

In this paper we investigate the localization properties of the bichromatic potential employed in current experiments, discussing the relationship with the discrete model in (\ref{eq:aa-hamiltonian}), and also the effect of the finite size of the system due to the presence of an additional harmonic confinement. We will provide evidences on how exponentially localized states emerge upon increasing the  degree of incommensurability of the system, and how the transition from extended to localized states can be extracted from the 
momentum distribution, that is easily accessible in current experiments from the absorption images after a time of flight \cite{roati}.

\section{The bichromatic lattice}
Recent experiments with cold atomic samples feature a one-dimensional bichromatic potential obtained by superimposing two optical lattices, of the form \cite{fallani,roati}
\begin{equation}
V_{b}(x)=s_1E_{R1}\sin^2(k_1x)+s_2E_{R2}\sin^2(k_2x + \phi)
\end{equation}
where $k_i = 2\pi/\lambda_i$ ($i=1,2$) are the lattice wavenumbers, $s_i$ are the heights of the two lattices in units of their recoil energies $E_{Ri} = h^2/(2m\lambda_i^2)$, and $\phi$ is an arbitrary phase.  In general, the potential of wavelength $\lambda_1$ is used to create a tight binding primary lattice, of period $d=\lambda_1/2$, that is weakly perturbed by the secondary lattice of wavelength $\lambda_2$. 

The single particle properties of the potential $V_b(x)$ are obtained by diagonalizing the hamiltonian
\begin{equation}
H=-\frac{\hbar^2}{2 m}\nabla^2_x +V_{b}(x),
\end{equation}
that can be conveniently rewritten in units of main lattice parameters ($k_{1}^{-1}$ and $E_{R1}$ as units of lenghts and energies, respectively) as
\begin{eqnarray}
{H}/{E_{R1}}&=&-\nabla^2_{\xi} +s_1\sin^2(\xi)+s_2\beta^2\sin^2(\beta\xi+\phi)\nonumber\\
&\equiv& H_1 + s_2\beta^2\sin^2(\beta\xi+\phi)
\label{eq:dimesionless-bichromatic}
\end{eqnarray}
with $\xi=k_{1} x$, $\beta=\lambda_1/\lambda_2$ and $E_{R2}/E_{R1}=\beta^2$.

It can be easily demonstrated that the presence of the second lattice does not change substantially the position of the minima but instead shifts the energies in a range of size 
$s_2\beta^2$ \cite{guarrera}.

\subsection{The tight binding limit: mapping onto the Aubry-Andr\'e model}

In the tight-binding limit this system can be mapped to the Aubry-Andr\`e model 
\cite{aubry-andre} by expanding the particle wavefunction $\psi$
over a set of Wannier states, maximally localized at the minima of the primary lattice (from now on ``lattice sites'', labelled by $j$) $w_j=w(\xi-\xi_j)$, 
in the \textit{lowest band}, $|\psi\rangle=\sum_j c_j|w_j\rangle$. 
The Aubry-Andr\'e model can be written in terms of the projector over the Wannier states as 
\begin{equation}
H\longrightarrow \sum_{ij}|w_i\rangle\langle w_{i}|H|w_{j}\rangle\langle w_j|
\end{equation}
or in terms of the $c_j$ coefficients 
\begin{equation}
H\longrightarrow \langle \psi|H|\psi\rangle=\sum_{ij}c_i^*c_j\langle w_{i}|H|w_{j}\rangle.
\end{equation}

Let us evaluate explicitly the expectation value of the hamiltonian on the Wannier basis;
in the tight binding limit we can make the following approximation 
\begin{eqnarray}
\label{eq7}
\langle w_{i}|H|w_{j}\rangle &\approx& \epsilon_0\delta_{ij} -J\delta_{i,j\pm 1}+ \delta_{ij}s_2\beta^2\int d\xi~ \sin^2(\beta\xi  + \phi)|w_i(\xi)|^2
\end{eqnarray}
with
\begin{equation}
\epsilon_0=\int d\xi~w_{i}(\xi)H_{1}w_i(\xi)\;;
\qquad
J=\int d\xi~w_{i+1}(\xi)H_{1}w_i(\xi)
\label{eq:jtunnel}
\end{equation}
where we have retained only the onsite contribution of the secondary lattice \cite{guarrera, giamarchi} and the tunneling between nearest neighbouring sites \cite{jacksh}
(for shallow lattices, the interaction
between Wannier states beyond nearest neighbours becomes relevant too \cite{holthaus}).
The integral in last term of Eq. (\ref{eq7}) can be rewritten by using the relation 
$\sin^2(\beta\xi  + \phi)=(1-\cos(2\beta\xi  + 2\phi))/2$, and noticing that
\begin{eqnarray}
\int d\xi~ \cos(2\beta\xi +\phi')|w_i(\xi)|^2=
\cos(2\pi\beta i+\phi')\int d\xi~ \cos(2\beta\xi )|w(\xi)|^2.
\end{eqnarray}
Finally, by dropping constant terms and defining
\begin{equation}
\Delta=\frac{s_2\beta^2}{2}\int d\xi~ \cos(2\beta\xi )|w(\xi)|^2
\label{eq:delta}
\end{equation}
we have
\begin{equation}
\langle w_{i}|H|w_{j}\rangle \approx-J\delta_{i,j\pm 1}
- \delta_{ij}\Delta\cos(2\pi\beta i+\phi')
\end{equation}

Therefore, in the tight-binding approximation the hamiltonian for the bichromatic potential can be cast in the form of the Aubry-Andr\`e model \cite{aubry-andre,thouless}
\begin{equation}
H=-J\sum_{j} \left(c_{j+1}^*c_j + c_j^*c_{j+1}\right) + \Delta\sum_j\cos(2\pi \beta j+ \phi) c_j^*c_j
\label{eq:aubry-andre}
\end{equation}
or \cite{ingold}
\begin{eqnarray}
H &=& -J\sum_{j}\left(|w_j\rangle\langle w_{j+1}|+|w_{j+1}\rangle\langle w_j|\right)
+  \Delta\sum_j\cos(2\pi \beta j+ \phi') |w_j\rangle\langle w_j|
\label{eq:discrete_model}
\end{eqnarray}
where the characteristic parameters $J$ and $\Delta$ are connected to the parameters directly tunable in the experiment, $s_1$, $s_2$ and $\beta$, via Eqs. (\ref{eq:jtunnel}), (\ref{eq:delta}). These can be estimate by using a Gaussian approximation for the Wannier functions, $|w(\xi)|^2\simeq (1/\sqrt{\pi})s_1^{1/4}\exp(-\sqrt{s_1}\xi^2)$, that gives $J= (4/\sqrt{\pi})s_1^{0.75}\exp(-2\sqrt{s_1})$ \cite{zwerger} and  $\Delta=(s_2\beta^2/2)\exp(-\beta^2/\sqrt{s_1})$ \cite{holthaus}.  
A more precise expression can be obtained by calculating numerically the integrals in Eqs. (\ref{eq:jtunnel}), (\ref{eq:delta}) in terms of the maximally localized Wannier functions.
This yields \cite{gerbier}
\begin{equation}
J\simeq1.43 s_1^{0.98} \exp(-2.07 \sqrt{s_1})\;,
\end{equation}
while for $\Delta$ we can make the following ansatz
\begin{equation}
\Delta\simeq\frac{s_2\beta^2}{2}e^{\displaystyle-\beta^\alpha/{s_1}^{\gamma}}
\label{eq:delta-fit1}
\end{equation}
with $\alpha$ and $\gamma$ to be determined from a fit of the numerical evaluation of the integral in (\ref{eq:delta}).
We note that in the literature it is common to find $\Delta$ defined 
without the exponential factor  \cite{giamarchi,fallani,roati};
though this is a legitimate definition of the intensity of the secondary lattice, the contribution of the exponential correction may be necessary for a precise mapping between the continuous and the discrete models, depending on the values of $s_1$ and $\beta$.

\subsection{Tuning the degree of commensurability}

A common choice in the study of the Aubry-Andr\'e model is the inverse of the golden ratio, $\beta=\varphi^{-1}\equiv (\sqrt{5} - 1)/2$, for which the model displays a ``metal-insulator'' phase transition from extended to localized states at $\Delta/J=2$ \cite{aubry-andre}. In particular, the insulator phase is characterized by the absence of mobility edges, all the eigenstates being exponentially localized with the same localization length $l=1/\ln(\Delta/2J)$. This case has been extensively investigated in the literature, see e.g. \cite{thouless,kohmoto,ingold,aulbach}. 

From the mathematical point of view, a necessary and sufficient condition to display a sharp transition is $\beta$ to be an irrational Diophantine number \cite{diophantine}. However, 
in the real world one has to be aware of two facts: \textit{i)} real system have a finite size; \textit{ii)} experimentally (and numerically) the wavelenghts can be given with a finite number of digits (in practice all number are rational). This means that what really matters is not the distinction between commensurable and incommensurable, but the actual \textit{degree of commensurability} and the ratio between the periodicity of the potential and the system size (fixed by the boundary conditions). In practice, in order to observe a transition from extended to localized states it is sufficient to have a large enough number of lattice sites within the actual periodicity of the potential, and a system size that does not greatly exceed the latter (in order to avoid periodic replicas).

\begin{figure}
 \centerline{\includegraphics[height=0.55\columnwidth,angle=-90]{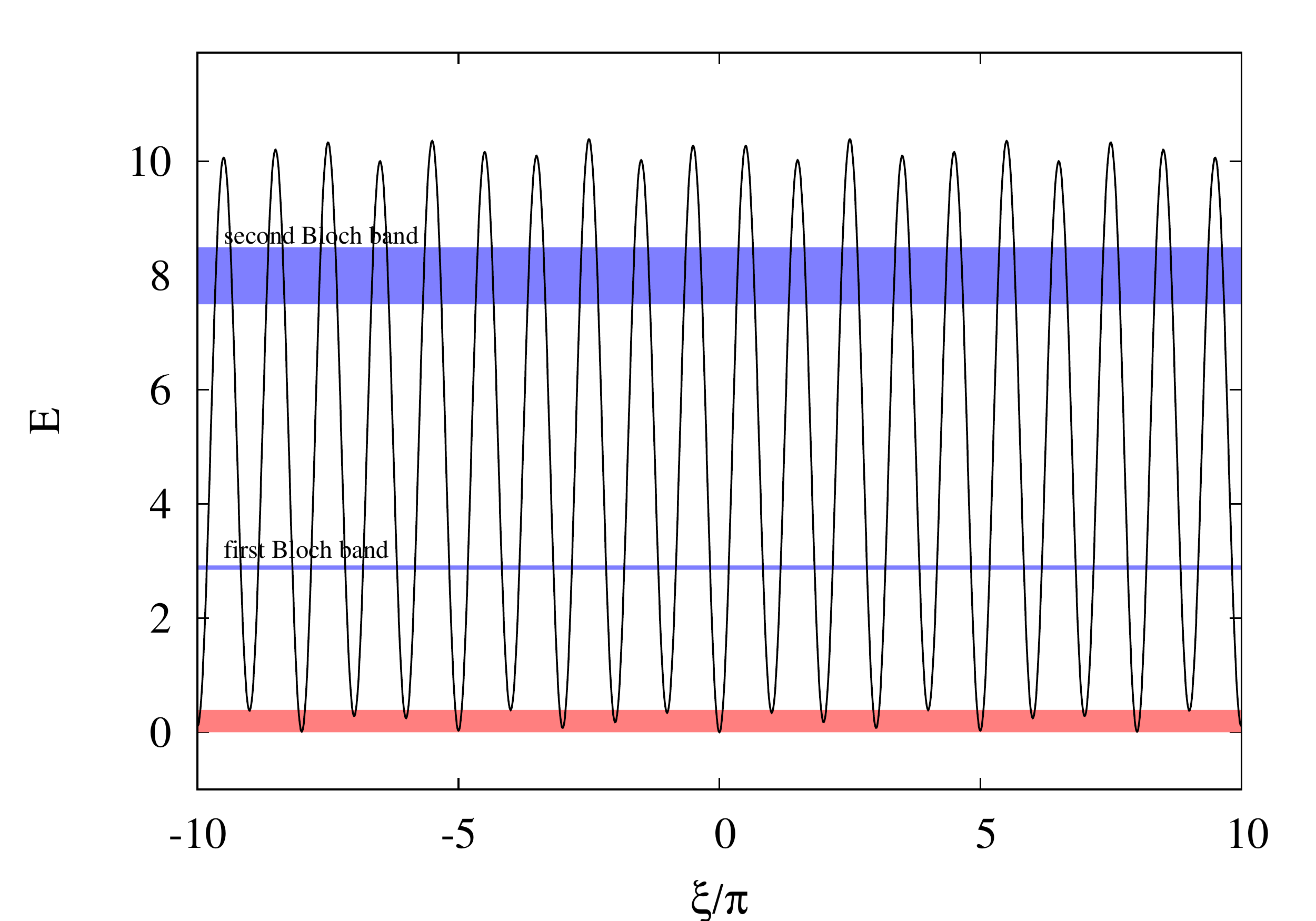}}
  \caption{Plot of the bichromatic potential for $s_1=10$ ($J\approx0.02$), $\Delta/J=10$.
  The blue stripes indicate the first and second Bloch bands of the primary lattice, whereas the red one represents the amplitude $2\Delta$ of the perturbation induced by the secondary lattice.  Notice that the latter is on an energy scale much lower that the bottom of the lowest band.}
 \label{fig:sketch}
\end{figure}

To illustrate how the properties of the system change by varying the degree of commensurability, we consider the solution of the full hamiltonian in Eq. (\ref{eq:dimesionless-bichromatic}), obtained by solving the Schr\"odinger equation (for convenience here we fix $\phi=0$) 
\begin{equation}
\left[-\nabla^2_{\xi} +s_1\sin^2(\xi)+s_2\beta^2\sin^2(\beta\xi)\right]\psi(\xi)=
E\psi(\xi).
\end{equation}
Here we focus on the tight-binding regime, and throughout this paper we fix $s_{1}=10$ ($J\approx0.02$). With this choice, an accurate expression for $\Delta$ is given by
\begin{equation}
\Delta\simeq \frac{s_2\beta^2}{2} e^{\displaystyle-a\beta^y}
\label{eq:delta-fit2}
\end{equation}
where the parameters $a$ and $y$ can be obtained by a numerical fit, yielding $a=0.385$, $y=1.95$.
In Fig. \ref{fig:sketch} we plot a sketch of the potential for $\Delta/J=10$ (corresponding to $s_{2}\approx0.47$), compared with the first and second Bloch band of the primary lattice. This picture shows that the large separation between the first and second bands indeed justifies the mapping onto the \textit{single band} Aubry-Andr\'e model, and that the perturbation introduced by the secondary lattice is at an energy scale much below the lowest band.

Differently from the common choice of periodic boundary conditions \cite{thouless,aulbach}, here we consider instead vanishing boundary conditions to account for finite size effects, in view of the fact that in the experiments there is always a residual harmonic confinement that introduces a natural cutoff at large distances (this choice does not affect the bulk properties). The specific effect of the harmonic confinement will be discussed later on.

\begin{figure}
 \centerline{\includegraphics[width=0.7\columnwidth]{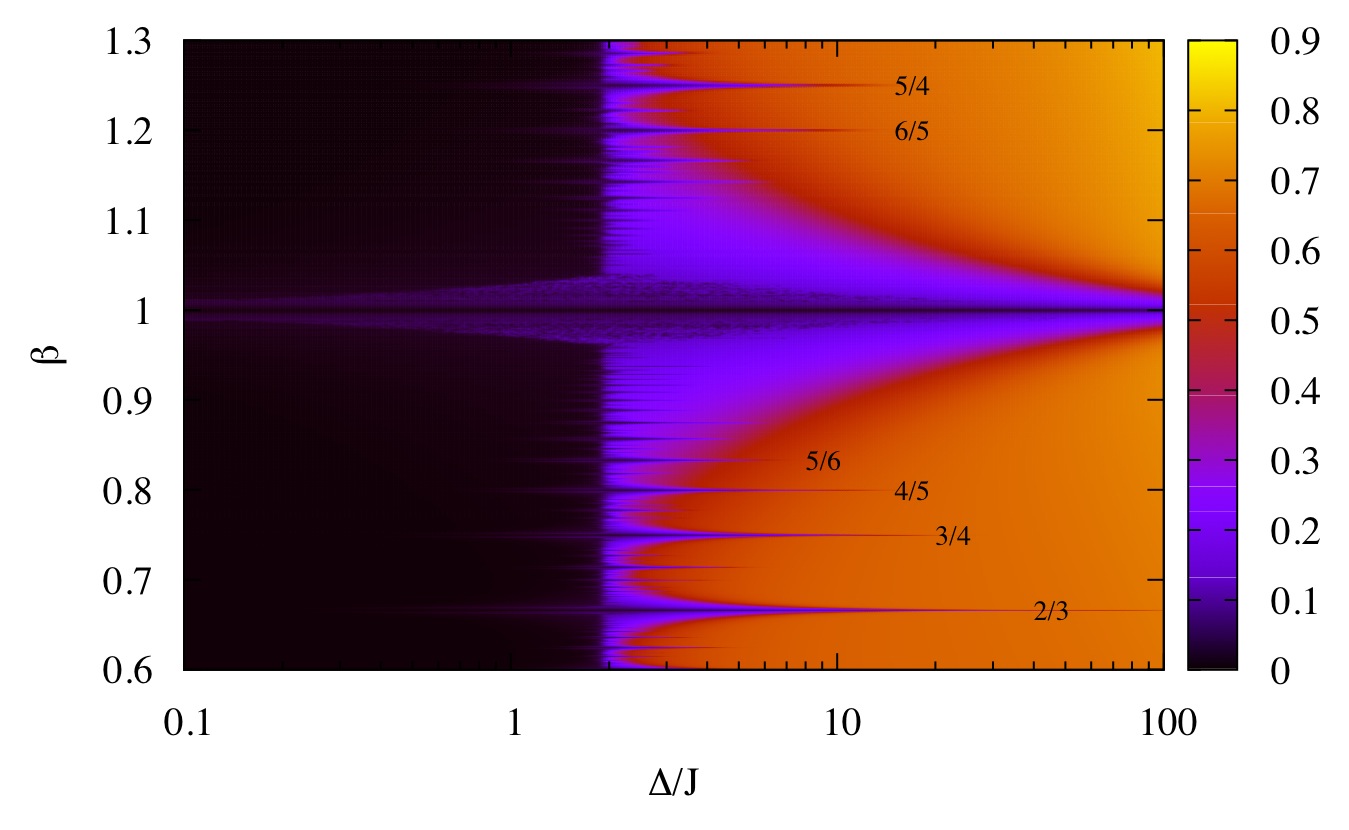}}
  \caption{Color plot of the inverse participation ratio ($P^{-1}$) for the bichromatic potential as a function of $\Delta/J$ and the ratio $\beta$ of the two wavelenghts. Black regions correspond to extended states ($P^{-1}\approx0$), whereas those colored in orange/yellow to localized states ($P^{-1}\gtrsim 0.5$). The size of the intermediate palette of colours measures the width of the crossover. The labels indicate the position of simple ratios of the two wavelenghts.  }
 \label{fig:ipr}
\end{figure}

To illustrate how the properties of the system change by varying $\beta$, in Fig. \ref{fig:ipr} we plot the inverse participation ratio $P^{-1}=\int d\xi |\psi(\xi)|^4$,
that measures the inverse of the number of occupied lattice sites \cite{ingold}. 
This picture shows that it is possible to have localization for a large enough $\Delta/J$, in a wide range of values of $\beta$, and that the transition broadens in the neighbourhood of simple rational values.

\subsection{A Fibonacci sequence}

To get further insight on how the localization transition changes when we increase the incommensurability of the system here we consider a set of bichromatic potentials whose wavelengths are in the ratio $\beta_{n}=f_{n}/f_{n+1}$ of two consecutive numbers of the Fibonacci sequence, defined by the recursion relation $f_{n+1}=f_n+f_{n-1}$ with $f_0=f_1=1$ (each number after the first two is the sum of the previous two numbers), $f_{n}=$1, 1, 2, 3, 5, 8, 13, 21, 34, 55, 89, 144, 233, etc. \cite{kohmoto}.
As an example here we choose $\Delta/J=10$, for which we expect localization in the incommensurate limit $\beta=\varphi^{-1}$ \cite{ingold}. 

\begin{figure}
 \begin{center}
 \includegraphics[width=0.35\columnwidth]{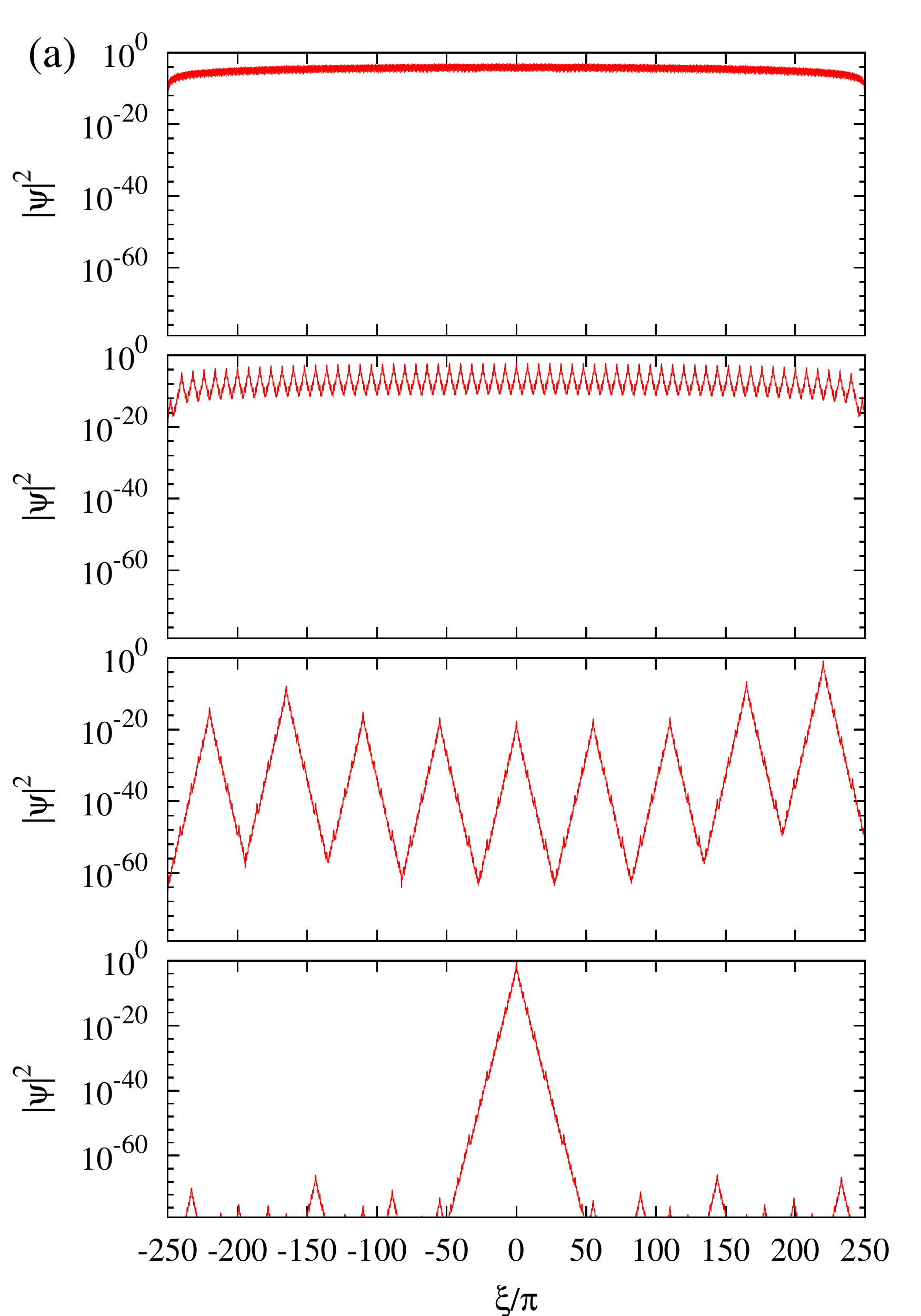}
 \includegraphics[width=0.35\columnwidth]{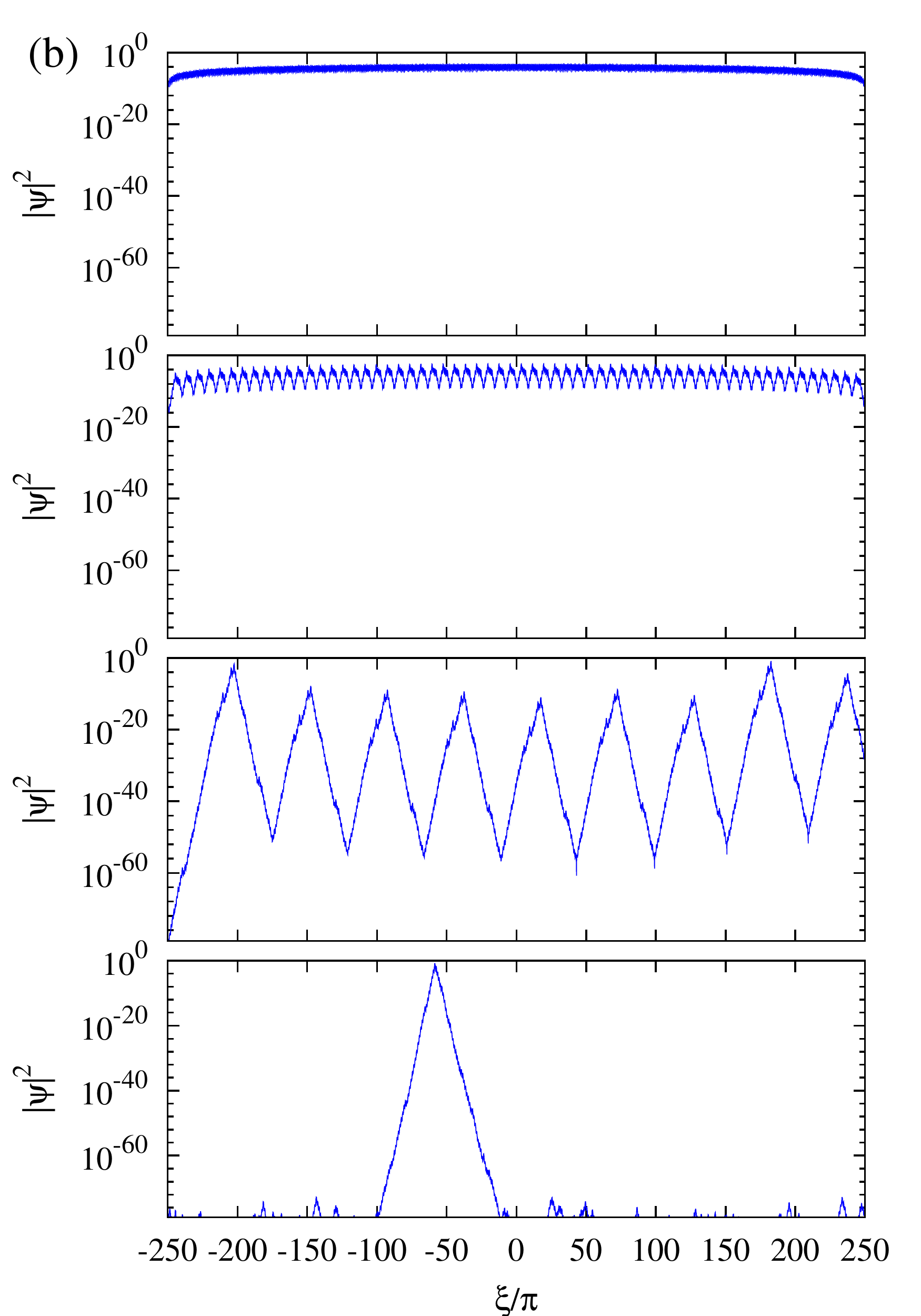}
\end{center}
 \caption{Plot the groundstate density for the bichromatic (a) and random (b, see \S\ref{sect:random}) potential
for $\beta_{0}= 1/1$, $\beta_{4}= 5/8$, $\beta_{8}= 34/55$, 
$\beta_{12}= 233/377$ (from top to bottom), and  $\Delta/J=10$, for a system of length $L=500$ (number of lattice sites).}
 \label{fig:fibonacci}
\end{figure}

In Fig. \ref{fig:fibonacci}a we plot the groundstate density of the bichromatic potential, for $\beta_{0}= 1/1$, $\beta_{4}= 5/8$, $\beta_{8}= 34/55$, 
$\beta_{12}= 233/377$, for a system of length $L=500$ (number of lattice sites) 
.
This picture shows how the density distribution of the groundstate gets modified as we increase the actual periodicity of the system, with peaks localized every $f_{n+1}$ sites of the main lattice, characterized by an exponential decay away from the localization center over an increasing number of lattice sites, $|\psi(\xi)|^2\approx \exp(-|\xi-\xi_0|/l)$. As the ratio $\beta$ of the two wavelengths approaches an irrational number the distance between two consecutive localized states goes to infinity, and in the limit only one state survives. 
Note that the localization is not due to the suppression of tunneling -- for $\beta$ rational the states are extended -- but to the randomization of onsite energies characteristic of Anderson localization \cite{grempel}.

\subsection{Quasiperiodic vs random}
\label{sect:random}

To compare the behaviour of the quasiperiodic lattice with that of a pure random system we consider a set of disordered lattice potentials, but  with a periodicity of $f_{n+1}$ sites as before, obtained by adding to the primary lattice a sequence of $f_{n+1}$ on-site energies 
$\epsilon_j$ extracted from a uniform random distribution in the range$[-\Delta,\Delta]$. 
This system is described by the tight binding hamiltonian\begin{equation}
H = -J\sum_{j}\left(|w_j\rangle\langle w_{j+1}|+|w_{j+1}\rangle\langle w_j|\right)
+ \sum_j \epsilon_j |w_j\rangle\langle w_j|
\end{equation} 
obtained by replacing the second term in Eq. (\ref{eq:aubry-andre}) by a random on-site energy term\footnote{In practice we have solved the continuum version of this model, 
obtained by shifting the minima of the primary lattice by $\epsilon_{j}$.}. 
In the limit $n\rightarrow\infty$ this system realizes the 
one-dimensional Anderson model \cite{anderson} with nearest-neighbour coupling. 
The corresponding groundstate distributions for $\Delta/J=10$ are shown in Fig. \ref{fig:fibonacci}b. 
This picture evidences a very similar behaviour of the two models,
suggesting that the mechanism of localization in quasiperiodic potentials is essentially the same of Anderson localization in disordered systems, irrespective of the fact
that the shuffling of the energies among different lattice sites occurs in a correlated or uncorrelated way (for the quasi-periodic and random cases, respectively) \cite{grempel}. This correspondence is remarkable, and supports the use of bichromatic potentials as effective tool for implementing quasi-disorder in current experiments with ultracold atoms \cite{damski,fallani,giamarchi,roati}.
The main difference is that for pure random system, in one (and two) dimension all states are always localized, for any amount of disorder \cite{ingold,kramer}. Instead, one-dimensional quasiperiodic potentials may show both extended and localized states, displaying a ``metal-insulator'' transition as a function of the strength of disorder \cite{aubry-andre,grempel}.

\section{Finite size systems: the effect of harmonic confinement}

The localization properties of a bichromatic potential have been recently demonstrated experimentally in \cite{roati}, where it has been reported the first observation of Anderson localization for a noninteracting matter wave. 
In this section we analyze the localization properties of this specific system, discussing the broadening of the ``metal-insulator" transition for the specific choice of the wavelenghts, the effect of the harmonic confinement, and the modification of the momentum distribution across the transition.

In the experiment \cite{roati} a noninteracting Bose-Einstein condensate of $^{39}$K atoms is initially prepared in a 3D harmonic potential $V_{ho}(x,r_\perp)=m\omega^2 x^2/2 + m\omega_\perp^2 r_\perp^2/2$, $r_\perp$ being the radial coordinate in the transverse direction (with respect to the lattice), and then loaded in a bichromatic potential of wavelengths 
$\lambda_1=1032$ and $\lambda_2=862$, yielding $\beta=1.1972\dots$. In the most commensurate hipotesys (neglecting decimal digits in the wavelenghts), this choice corresponds to a strict periodicity every $n_1=432   $ sites of the primary lattice.
Since the condensate is non-interacting, the problem is separable and the axial and radial wavefunctions factorize. In this case the presence of the bichromatic lattice affects only the axial wavefunction, that can be obtained by diagonalizing the hamiltonian
\begin{equation}
H=-\frac{\hbar^2}{2 m}\nabla^2_x +V_{b}(x)+ \frac{1}{2}m\omega^2 x^2.
\end{equation}
 
The only difference with what discussed in the previous sections is the presence of the axial harmonic confinement\footnote{In the numerical calculations we have also included a tiny phase shift between the bichromatic and harmonic potentials in order to avoid perfectly symmetric solutions, that may be quite rare in the experiment.}
, that fixes the size of the system to a value of the order of the oscillator length, $a_{ho}=\sqrt{\hbar/m\omega}$.  In the following we will assume $a_{ho}\ll n_1\lambda_1$ in order to consider the case of systems whose size does not exceed the overall periodicity of the bichromatic potential. Moreover, we may expect the localization properties to be not substantially affected by the harmonic trap when the system extends over several lattice sites, that is when $a_{ho}\gg d$. In this case the harmonic trap only affects the boundary conditions but not the localization behaviour in the bulk.
In the experiment the lowest value of the trapping frequency is fixed by the residual trapping due to the focusing of the laser beams that creates the primary lattice, $\omega_{res}\approx2\pi\times 0.5$ Hz. Instead, the measurements of the momentum distribution have been taken with $\omega\approx 2\pi\times 5$ Hz. Both these values satisfy the above requirements, and correspond to $a_{ho}/d=44,14$ respectively.

\begin{figure}
 \begin{center}
 \includegraphics[width=0.8\columnwidth]{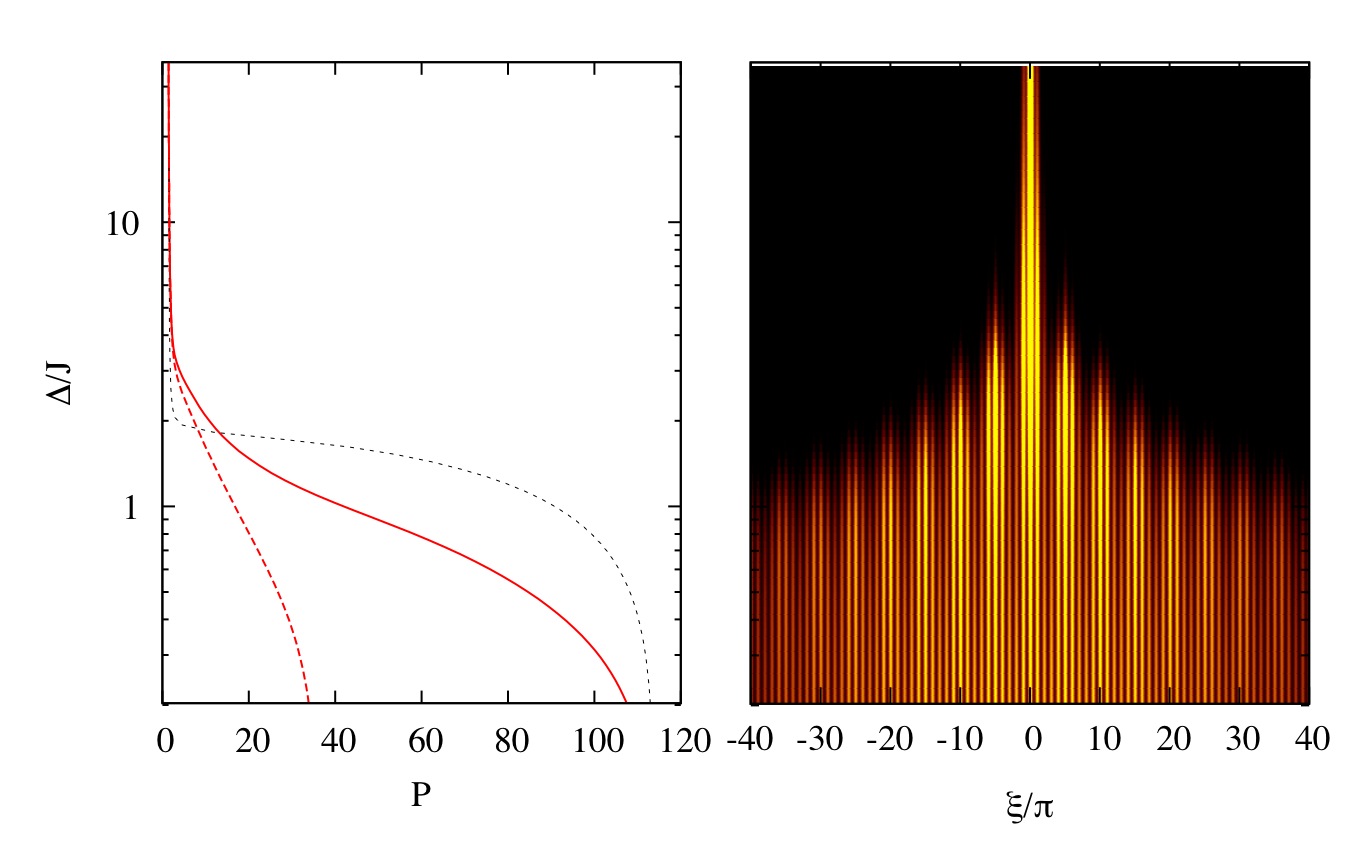}
\end{center}
 \caption{Left: participation ratio of the groundstate of the bichromatic potential of Ref. \cite{roati} as a function of $\Delta/J$, for a trapping of frequency $\omega_{res}=2\pi\times 0.5$ Hz (red, continuous line) and $\omega=2\pi\times 5$ Hz (red, dashed line). 
 The dotted line represents the participation ratio for $\beta=(\sqrt{5}-1)/2$, in the case $\omega=\omega_{res}$. 
Right: density plot of the groundstate for $\omega=\omega_{res}$.}
 \label{fig:plot05}
\end{figure}

In Fig. \ref{fig:plot05} we show  the participation ratio and the density plot of the groundstate of the bichromatic potential as a function of $\Delta/J$, for the residual trapping, $\omega=\omega_{res}$. We note that in \cite{roati} the intensity of disorder has been defined according to the common choice $\Delta={s_2\beta^2}/{2}$ \cite{giamarchi,fallani}; here we will instead use the definition in Eq. (\ref{eq:delta-fit2}) for a more precise comparison with the discrete Aubry-Andr\'e model.
This picture shows that for the particular choice of the wavelenghts of \cite{roati} there is a broadening of the ``metal-insulator'' transition with respect to the ``incommensurate'' case   $\beta=\varphi^{-1}$, and that the eigenstate size becomes comparable with the lattice spacing at about $\Delta/J=5$.

\begin{figure}[t]
 \begin{center}
 \includegraphics[width=0.4\columnwidth]{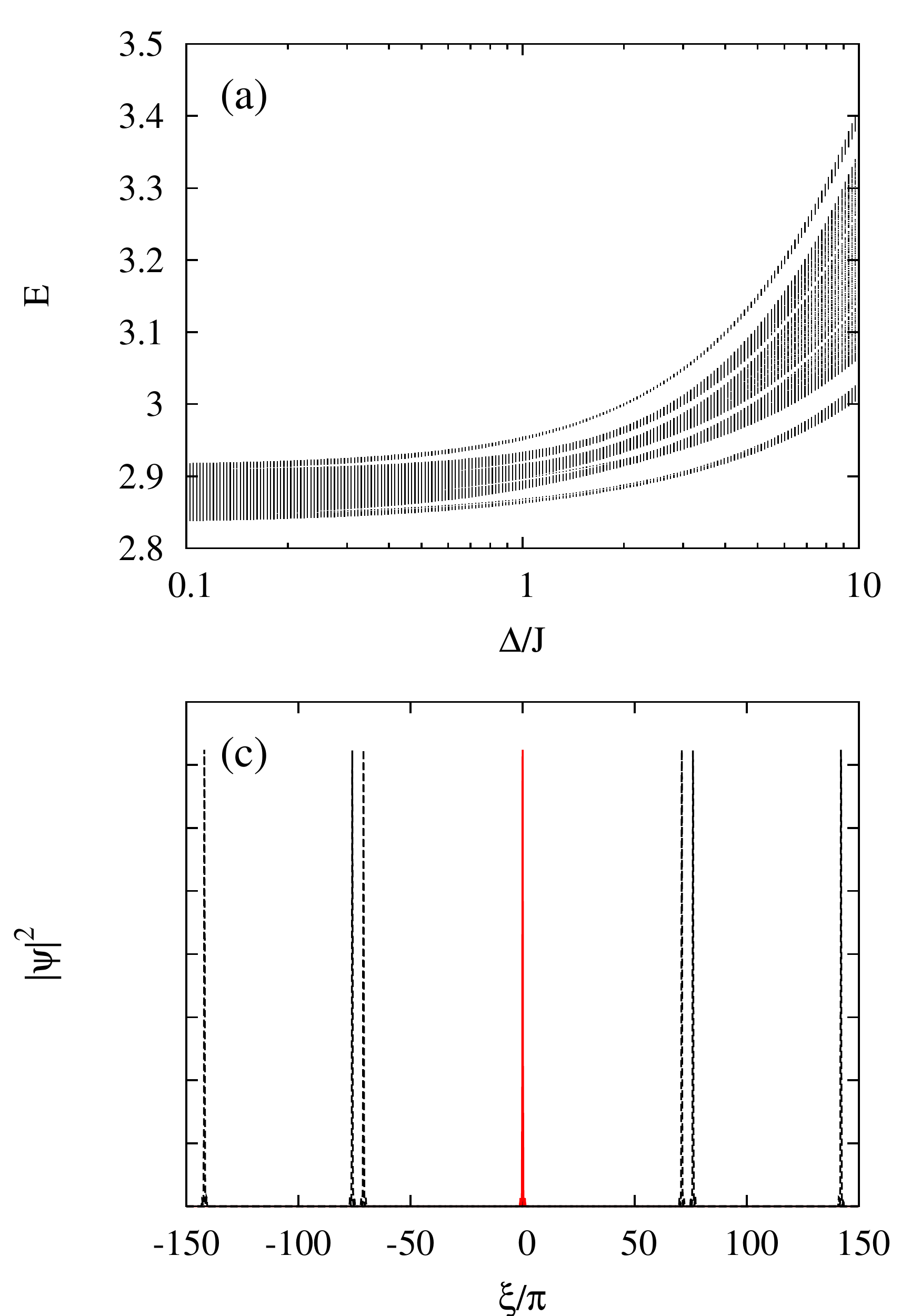}
 \includegraphics[width=0.4\columnwidth]{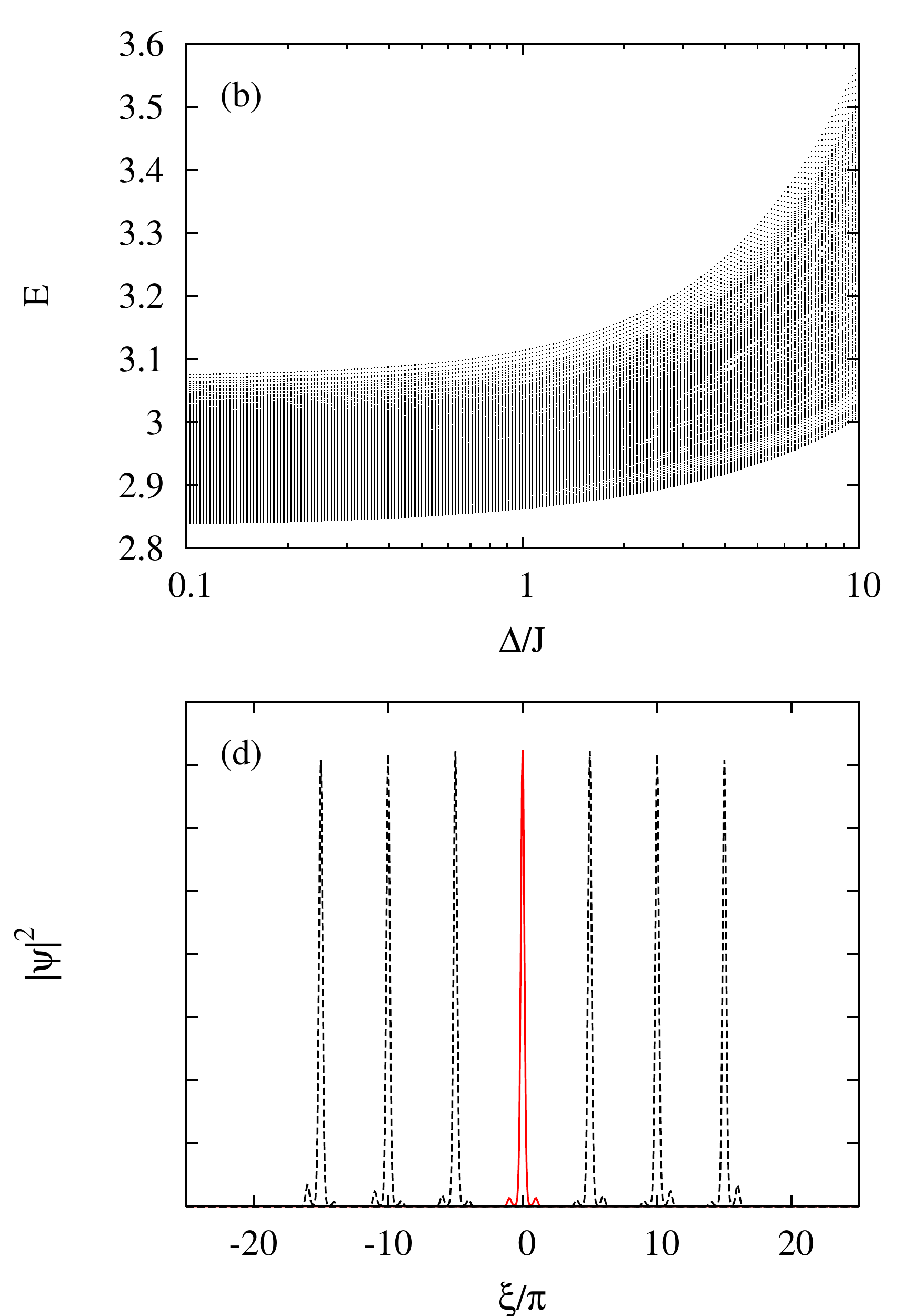}
\end{center}
 \caption{Top row: spectrum of the bichromatic potential as a function of $\Delta/J$, for $\omega=2\pi\times 0.5$ Hz (a) and $\omega=2\pi\times 5$ Hz (b). 
(c),(d): the corresponding ground states (at the center, red continuous line) and the first six excited states (dashed lines) for $\Delta/J=10$. }
 \label{fig:spectrum-ho}
\end{figure}

The modification of the spectrum as a function of $\Delta/J$ is shown in Fig. \ref{fig:spectrum-ho}ab for both values of the trapping frequency.
Here we have considered the energy range corresponding to the lowest band of the primary lattice. This picture illustrates that, in case of weak confinement ($\omega=2\pi\times 0.5$ Hz), several mini gaps open in the spectrum as we turn on the secondary lattice, the main ones being those corresponding to the beating of the two lattices \cite{lye}. Instead, for a tighter trapping ($\omega=2\pi\times 5$ Hz) the spectrum is 
deeply affected by the presence of the harmonic potential, that produces an enlargement of the  band size and a different redistribution of the energies.
In particular, the presence of the harmonic confinement affects the spatial arrangement of the localized eigenstates, as shown in Fig. \ref{fig:spectrum-ho}cd. For a shallow enough confinement the localized eigenstates are shuffled all along the system, whereas a tighter confinement produces a hierarchical arrangement starting from the trap center, with eigenstates neighboring in energy that are disposed every five sites (with a reflection symmetry about the trap center).
This feature allows to tune the arrangement of the eigenstates around the trap center by changing the confinement frequency, without affecting their exponential localization behavior \cite{roati}.

\subsection{Momentum distribution}

\begin{figure}[t]
 \begin{center}
 \includegraphics[width=0.35\columnwidth]{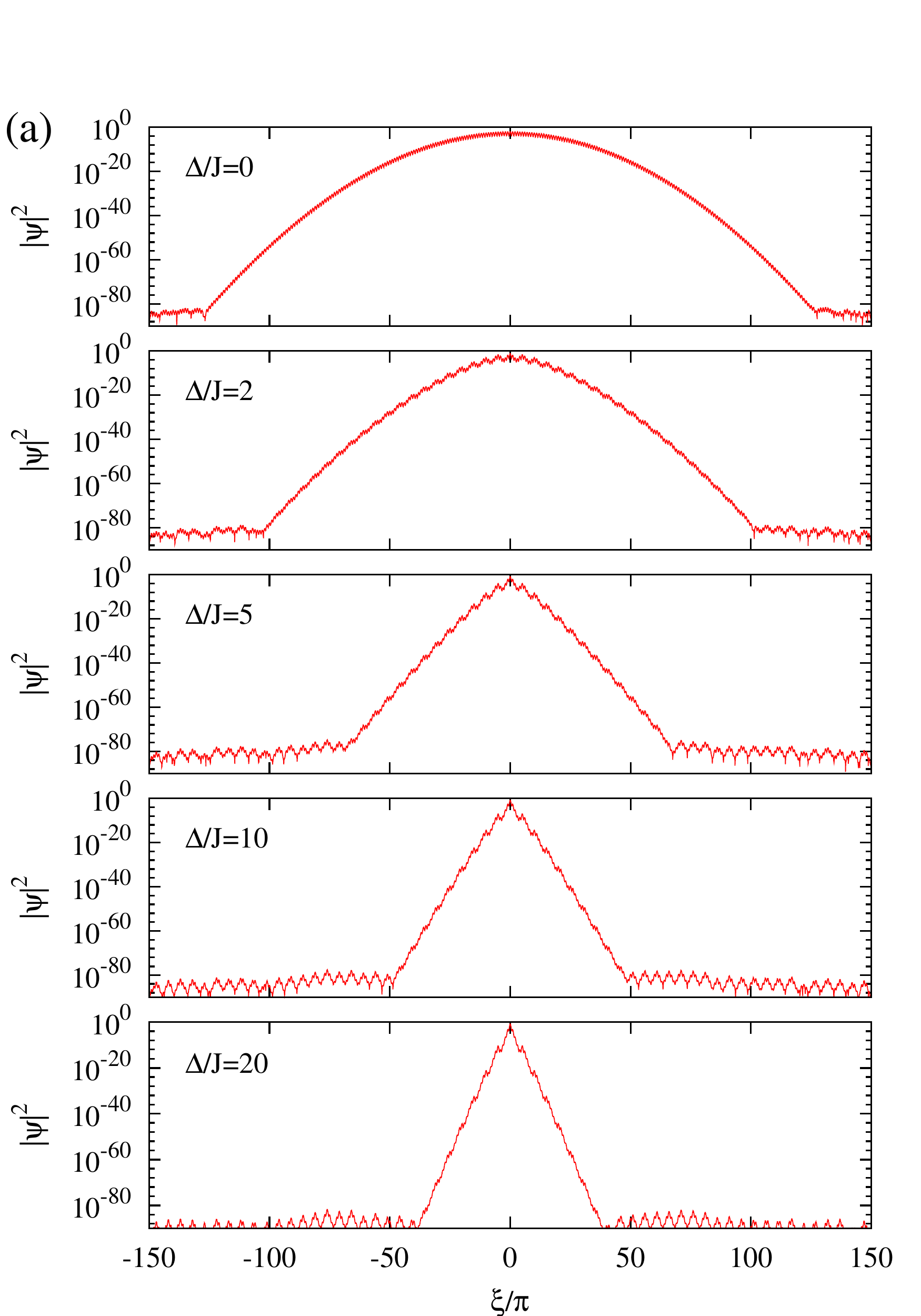}
 \includegraphics[width=0.35\columnwidth]{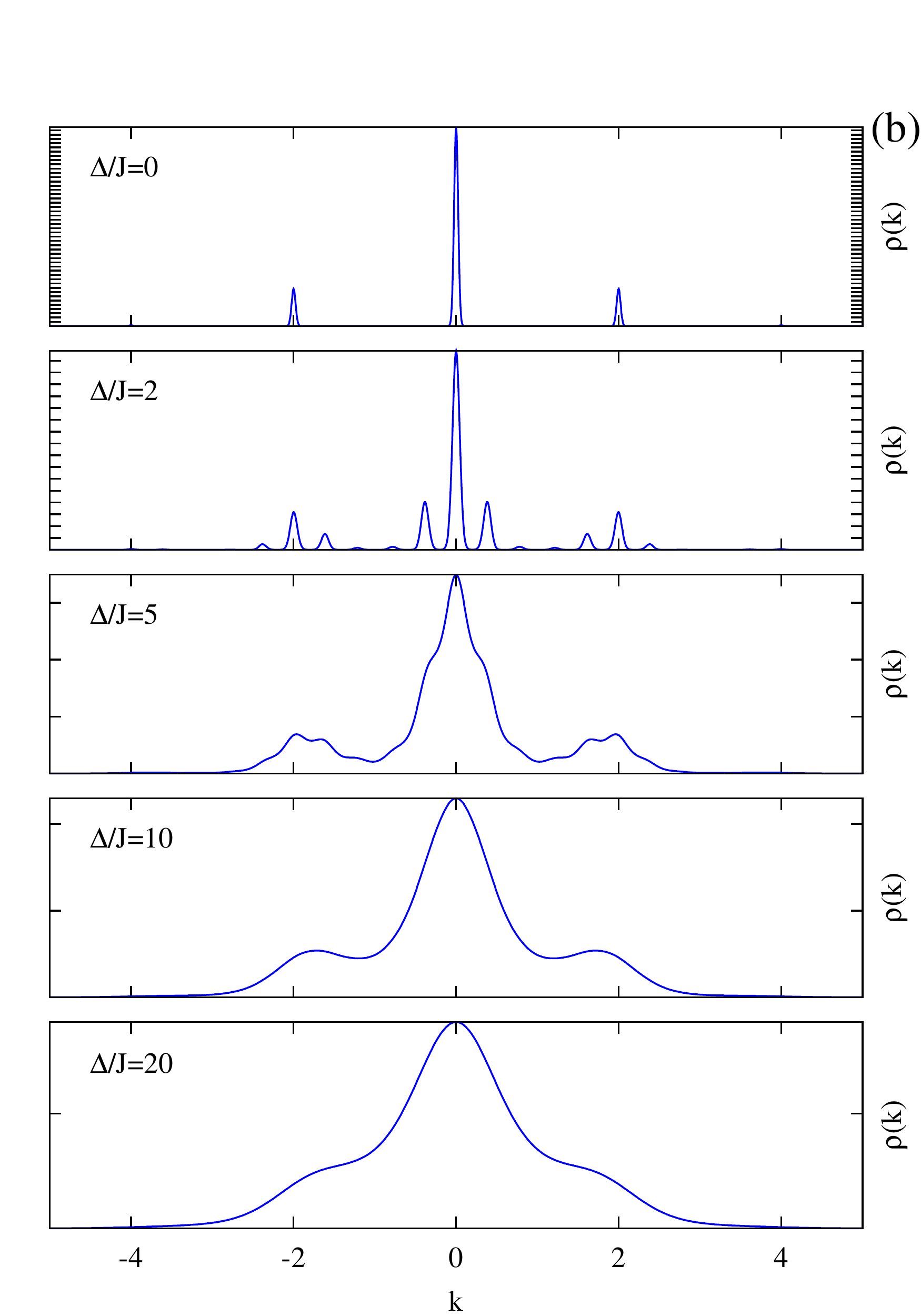}
\end{center}
 \caption{Plot of the groundstate density (left, in log scale) and of its momentum distribution (right, note the different scale of the vertical axis) for different amplitudes of the secondary lattice, $\Delta/J=0, 2, 5, 10, 20$ from top to bottom, in the presence of a harmonic trapping of frequency $\omega=2\pi\times 5$ Hz. }
 \label{fig:ho}
\end{figure}

Important indications about the modification induced by the secondary lattice on the eigenstates of the system can be obtained from their momentum distribution. This is shown in Fig. \ref{fig:ho} where we plot the groundstate density $|\psi(\xi)|^2$ (left) and its momentum distribution $\rho(k)=|\tilde{\psi}(k)|^2$ (right) for increasing values of $\Delta/J$, with $\tilde{\psi}(k)=\int d\xi \psi(\xi) e^{{ik\xi}}$. Here we consider the same value of the trapping frequency $\omega=2\pi\times 5$ Hz used in \cite{roati} to make the analysis of the momentum distribution after the time-of-flight.

When only the primary lattice is present, $\rho(k)$ displays the typical interference pattern with three peaks at $k=0,\pm2k_1$ reflecting the periodicity of the system and the fact the the eigenstates are extended. The tiny width of the peaks indicates that the wavefunction spreads over many lattice sites \cite{pedri}. 
When the secondary lattice is added, additional momentum peaks appear, corresponding to the beating of the two lattices. 
The presence of these peaks can be easily explained for a commensurate system defined by $n_{1}\lambda_{1}=n_{2}\lambda_{2}$ (here we consider $n_{2}>n_{1}\equiv L$ as in the experiment, $\beta=n_{2}/n_{1}$), by transforming the discrete model
in (\ref{eq:discrete_model}) in the momentum representation ($k_{l}=2 l/L$ in units of $k_{1}$)
\begin{equation} 
\tilde{H} = -2J \sum_{l=1}^{L} \cos(2\pi l/L)|k_{l}\rangle\langle k_{l}| 
+ \Delta\sum_{ll'} \tilde{\delta}_{l-l'} |k_{l}\rangle\langle k_{l'}|
\end{equation}
where we have defined 
$|k_{l}\rangle=(1/\sqrt{L})\sum_{j}e^{i\pi k_{l}j} |w_{j}\rangle$, and
 \begin{equation}
\tilde{\delta}_{l-l'}\equiv
\frac{1}{2L}\sum_{j=1}^{L}\sum_{\pm}e^{-i2\pi j((l-l')/L\pm\beta)}=
\frac{1}{2L}\sum_{j=1}^{L}\sum_{\pm}e^{-i\pi(k_{l}-k_{l'}\pm 2k_{2})j}
\end{equation}
with $k_{2}=\beta k_{1}=\beta$. It is easy to prove that the term $\tilde{\delta}_{l-l'}$, that determines the coupling between different momentum components, 
is non vanishing when $|k_{l}-k_{l'}|=2 k_{2}-2k_{1}$ or $|k_{l}-k_{l'}|=4k_{1}-2 k_{2}$.
When going back to the continuous model, the above relations correspond to the additional peaks in the momentum distribution, that appear at a distance 
$\pm2(k_2 - k_1)$ around those at integer multiples of $\pm2k_{1}$ due to the primary lattice. 

By further increasing the intensity of the secondary lattice, the momentum distribution $\rho(k)$ broadens and its width eventually becomes comparable with that of the first Brillouin zone.
This take place in correspondence of  the appearance of exponentially localized states whose extension shrinks below that of a single lattice site. 
The residual modulation of the momentum distribution in the two bottom profiles of Fig. \ref{fig:ho} (right) indicates that the localization is nontrivial, in the sense that the tails of the eigenstates extend over several lattice sites even for large $\Delta/J$ (as shown in the left column). 
This analysis accounts for the behavior observed in \cite{roati} across the localization transition.

\section{Conclusions}

We have analyzed the single particle localization properties of a one-dimensional bichromatic potential obtained by superimposing two optical lattices of different wavelenghts, discussing how the degree of commensurability of the two lattices affects the appearance of exponentially localized states,  in the tight binding regime. We have shown that in order to observe a transition from extended to localized states in a finite size system what really matters is not the distinction between commensurable and incommensurable, but the possibility to have a large enough number of lattice sites within the actual periodicity of the potential.

We have also reviewed the mapping onto the discrete Aubry-Andr\'e model \cite{aubry-andre} and made a comparison with pure random disorder, providing evidence of the similarity of the localization mechanism in quasiperiodic and random systems, in support of the use of bichromatic potentials as effective tool for implementing quasi-disorder in current experiments with ultracold atoms \cite{damski,fallani,giamarchi,roati}.

Finally, we have discussed the effect of an additional harmonic potential and analyzed the modification of the momentum distribution across the ``metal-insulator'' transition,
 making a direct discussion of the experiment \cite{roati}, where it has been reported the first observation of Anderson localization of a noninteracting Bose-Einstein condensate in a quasiperiodic optical lattice.

\ack
This work has been stimulated by the many discussions with all the authors of \cite{roati}.
I am grateful to C. Fort, G. Roati and G. Modugno for the careful reading of the manuscript and useful suggestions, and to M. Larcher and F. Dalfovo for interesting discussions.

\section*{References}

\end{document}